\documentclass[12pt]{article}
\newcommand{\tp}{\otimes}             % "tensor product"
\newcommand{\ww}{\wedge}             % "tensor product"
\newcommand{\la}{\langle}  
\newcommand{\ra}{\rangle}  

\newcommand{\ppp}{\partial}
\newcommand{\be}{\begin{eqnarray}}
\newcommand{\ee}{\end{eqnarray}}
\newcommand{\bes}{\begin{eqnarray*}}
\newcommand{\ees}{\end{eqnarray*}}

\begin{document}
\title{Poincare Cartan Form for Gauge Fields in Curved Background.}
\author{Pankaj Sharan\\Physics Department, Jamia Millia Islamia, \\New Delhi 110 025, INDIA}
\maketitle

\abstract{The `directly Hamiltonian' field theory in the extended phase space is applied to 
gauge fields in curved spacetime  background. These fields being differential 
1-forms, have canonical momenta
which are 2-forms. The Poincare-Cartan 4-forms for matter and gauge fields
 have to be modified  with
the exterior derivatives replaced by the covariant derivative for 
maintaining gauge invariance.}

\section{Introduction}
In a recent communication \cite{sharan1} (called I here) 
a Hamiltonian formalism is developed in which 
scalar fields are associated with canonical momenta which are differential 1-forms. This
is a `directly Hamiltonian' formalism where canonical momenta are not defined through a
Lagrangian via time derivatives of fields. 
Rather, manifest invariance in arbitrary curved spacetime is maintained
at each step. The main features of this formalism are as follows. For 
background, motivation and other details see I.
\begin{enumerate}

\item 
If we treat all the four derivatives $\ppp_\mu\phi$ of the field
$\phi$ as one quantity then it follows we
should allow four components in momentum $p=p_\mu dx^\mu$ to be associated with one 
field variable $\phi$. Thus canonical momentum is a differential 1-form 
for a field which is a form of degree zero.
The `coordinate' $\phi$  and its `canonical momentum' $p$ are both differential forms but with 
the momenta being forms of a degree higher.

\item
The central object in Hamiltonian mechanics is the Poincare-Cartan (PC) differential 1-form
$pdq-Hdt$. The PC-form in field theory is a differential 4-form constructed
in the extended phase space. The standard structure of PC-form has a `fundamental'
part $pdq$ which is linear in the differential of the coordinates. This part
determines the relation of canonical momentum to gradient of the field.
The remaining part of PC-form is the Hamiltonian which is constructed out of 
fields and momenta. For a single free scalar field, for example,
we can write
\bes \Theta=(*p)\ww d\phi-H, \qquad H=\frac{1}{2}(*p)\ww p+\frac{1}{2}m^2\phi^2(*1)\ees
where the Hodge dual star operator associated with the Riemannian metric $g_{\mu\nu}$ of
spacetime is used to convert 1-form $p$ into 3-form $*p$ and 0-form (here constant 1)
 into volume 4-form $(*1)$. See the appendix A for notation.
\item In general the Hamiltonian $H$ is constructed as a 4-form (which is equivalent
to a scalar) with $\phi$ and $p$ just like a Lagrangian is constructed from $\phi$ and
$d\phi$ in Lagrangian field theory.
\item
The extended phase space has the structure of a fiber bundle where spacetime 
is the base manifold 
and the fiber consists of fields and momenta.

\item
The allowed field configurations are sections (i.e. mappings from the spacetime 
into the bundle) which are submanifolds of the bundle characteristic of the 5-form
$\Xi=-d\Theta$. This means that  tangent vectors of the  submanifold  annihilate  $\Xi$.

\item Observables of the theory are differential 4-forms integrated over spacetime.
For example,  $\phi(x)$ is not an observable, but a quantity of type
$\int \phi(x)j(x)(*1)$ is, where
$j(x)$ is a scalar `test function' which can be chosen to be non-zero in some small region
of spacetime. There is no Poisson bracket in the usual sense because coordinates and
momenta are forms of different degrees in this formalism. 
Instead, we use a generalization of the idea that
Poisson bracket between two observables determines rate of change of one observable when 
the other observable is the Hamiltonian or part of the Hamiltonian.
The generalized bracket was defined by Peierls
in the early days of quantum
field theory. 
\item Even in classical mechanics coordinates are zero-forms
and momenta are 1-forms. But the base manifold in classical mechanics is one
dimensional, and 1-forms in 1-dimension are as good as scalars or 0-forms.
Thus coordinates and momenta seem to be on the same footing and the Poisson bracket has
a coordinate-momentum symmetry. The Poisson brackets turn out to be equal-time
Peierls bracket in classical mechanics as has been shown by the author \cite{sharan2}.

\item The symmetries and conserved quantities can be discussed 
very easily in this formalism.
Invariance of the  PC-form, expressed as the vanishing Lie derivative $L_Y\Theta=0$, where  
$Y$ is the vector field of an infinitesimal symmetry transformation.
Noether's theorem holds
and the integral of the 3-form $i(Y)\Theta$ over a closed 3-surface is zero.
Usually, the three surface is taken made up of two space-like surfaces
and a cylindrical surface at infinity. The conservation
law takes its usual form in this setting.
\end{enumerate}

The purpose of the present paper is to extend this formalism for gauge fields.
In order to preserve gauge invariance we have to modify the structure of PC-form
from $*p\ww d\phi-H_\phi$ to $\Theta_\phi=*p\ww (d\phi+A\phi)-H_\phi$ 
where the 1-form $A$ is the gauge connection.
In order to include the gauge fields as dynamical fields we must define the PC-form
for $A$ as well. Since $A$ is already a 1-form its canonical momentum
related linearly to velocities $dA$  would be a 2-form.
Calling that $F$, the PC-form should be $\Theta_A=*F\ww(dA+A\ww A)-H_A$ 
as the natural gauge invariant object.

In the next section we introduce external gauge connection fields $A$ 
interacting with real scalar fields which have a an $O(N)$ gauge group.
We choose scalar fields for simplicity
to illustrate the idea. Spinors  involve tetrad fields
and will be  treated separately.
 In section 3 we develop the
formalism for the gauge field alone. The canonical momentum associated to connection field
$A$ turns out to be the curvature $dA+A\ww A$ on allowed characteristic submanifolds.
Finally in section 4 interacting fields $\phi$ and $A$ are considered and equations of motion derived. The issue of gauge fixing and Peierls bracket in this formalism will be treated in a separate communication.

\section{Scalar fields in an external gauge field}

If there are $N$  real scalar fields then the PC-form for them is 
\be \Theta_\phi = *p_i\ww d\phi^i-H \ee
where $H$ is a 4-form depending on all the fields and their momenta. We will omit the
index $i$ and write the form simply as $*p\ww d\phi-H$. If there is an 
internal space global $O(N)$ rotation (acting on the right) 
$\phi\to R^{-1}\phi$ we define 
the momenta to transform contra-gradiently $p\to R^Tp=pR$. 
This keeps the PC-form unchanged provided the covariant 
Hamiltonian $H$ remains unchanged too.

However, if the internal space rotation is `gauged', that is, if matrices $R$ depend
on spacetime coordinates $t^\mu$ then the PC-form must be modified with the introduction
of a gauge connection form $A$ which are new 1-form fields taking values as Lie
algebra elements.

We suggest the following expression for the PC-form
\be \Theta_\phi=*p\ww (d\phi+A\phi) -H(\phi,p). \ee
Here $A$ should be regarded as a matrix valued 1-form
\be A =A^a\tau_a= A^a_{\mu}dt^\mu \tau_a.\ee
The matrices $\tau_a$ are `generators' (that is Lie algebra basis elements)
of the internal symmetry group in the (real) representation to 
which fields $\phi_i$ belong. We assume
\be [\tau_a,\tau_b]={C_{ab}}^c\tau_c \ee 
where ${C_{ab}}^c$ are structure constants of the Lie group.
The Lie algebra in
the {\em adjoint} representation, which we need later, is defined by
\be {(T_a)_b}^c=-{C_{ab}^c}. \ee
It satisfies
\be {\rm Tr}[T_aT_b]=k_{ab}\ee
where constants $k_{ab}$ allow us to define an inner product on the Lie algebra. It is
well-known that $k_{ab}$ is non-degenerate for semi-simple Lie groups  and
moreover it is negative definite (as here) for compact groups.
We call its inverse by $k^{ab}$ where $k_{ab}k^{bc}=\delta_a^c$.

It clear that the PC-form is invariant if along with $\phi\to R^{-1}\phi$,
the connection transforms as
\be A\to R^{-1}AR+R^{-1}dR.\ee
The form  $-d\Theta$ is
\be -d\Theta &=&-(d*p)\ww(d\phi+A\phi)+(*p)\ww d(A\phi)+dH_\phi \nonumber \\
&=&-(d*p)\ww(d\phi+A\phi)+(*p)\ww [dA\phi-A\ww d\phi]+dH_\phi\nonumber \\
&=&-(d*p)\ww(d\phi+A\phi)-(*p)\ww A\ww d\phi+dH_\phi\nonumber \\
&=&-(d*p)\ww(d\phi+A\phi)-(*p)\ww A\ww(d\phi+A\phi)+dH_\phi\nonumber \\
&=&-(d*p+*p\ww A)\ww(d\phi+A\phi)+dH_\phi
\ee
where in the third step {\em we have omitted the term} $(*p)\ww dA\phi$ because it has
five factors of $dt^\mu$, three for $*p$ and two for 
\bes dA=A^a_{\mu,\nu}dt^\nu\ww dt^\mu \tau_a .\ees
Here the connection field $A$ is supposed to be given as an explicit 
function of spacetime
coordinates. Later, when we involve $A$ as a dynamical field we would be required to keep
this term because then 
\bes dA=dA^a_{\mu}\ww dt^\mu \tau_a \ees
would have only one factor of $dt^\mu$.

For a Hamiltonian of the form
\be H_\phi=\frac{1}{2}(*p)\ww p+\frac{1}{2}m^2\phi^T\phi(*1) \ee
the exterior derivative is 
\be dH_\phi &=& (d*p)\ww p+m^2\phi\,(*1)\ww d\phi \nonumber \\
&=& (d*p)\ww p+m^2\phi\,(*1)\ww (d\phi+A\phi-p)\nonumber \\
&=& (d*p+*p\ww A)\ww p+m^2\phi\,(*1)\ww (d\phi+A\phi-p)
\ee
where we can harmlessly add $A\phi-p$ to $d\phi$ because these terms
are 1-forms proportional to $dt$ which when multiplied to $(*1)$ 
gives zero because of five $dt$ factors.
Similarly, keeping an eye on eqn. (8), we can add $*p\ww A\ww p=0$ 
in the last step above because
it has five factors of type $dt^\mu$ which are zero in 4-dimensional spacetime. 
Substituting this expression for $dH_\phi$ into (8) we obtain
\be \Xi=-d\Theta = -(d*p+*p\ww A-m^2\phi (*1))\ww(d\phi+A\phi-p)\ee
The factorization of $\Xi=-d\Theta$ allows us a quick route to the field equations.
The second factor gives the definition of canonical momentum
in terms of velocities
\be p = d\phi+A\phi,\ee 
and the first factor has terms
\be d*p+*p\ww A &=&
d*(d\phi+A\phi)+*(d\phi+A\phi)\ww A \nonumber \\ 
&=& d*(d\phi+A\phi)-A\ww *(d\phi+A\phi)\nonumber \\
&=& (d-A)\ww *(d+A)\phi.
\ee
We have a slight abuse of notation here, the term $d\ww (\dots)$
is just $d(\dots)$.
This leads to the equation of motion for the field
\be [(d-A)\ww *(d+A)-m^2]\phi=0 \ee

\section{PC-form for the gauge field alone}

In the analysis of the last section the connection field is
given externally. To include the field into dynamics, we must write appropriate terms
for it in the PC-form. We first write the PC-form just for the connection field.

The connection potential $A$ is already a 1-form matrix which depends on
which representation we are considering. In the last section the fields $A$
were in the representation provided by matrices $\tau_a$ suitable for acting on the
matter fields whereas in this section matrices are in
the adjoint representation ${(T_a)_b}^c=-C_{ab}^c$ for the generators. 
Strictly, we should call the gauge potential in the
two cases by different symbols. But we use the same symbol for simplicity
hoping this to be kept in mind.  

To express 
the first term of the fundamental 4-form $\Theta_A$ with structure 
$*(\dots)\ww dA$ we must define  
the (star operated) canonical momentum, call it $*F$, a 2-form matrix
because $dA$ is already a 2-form.
Therefore $F$ is a 2-form matrix as well. 

The gauge transformations of $A$ is well-known $A\to R^{-1}AR+R^{-1}dR$. 
$R$ is the group matrix 
in the adjoint representation here.

We also know that
\bes dA+A\ww A 
&\to& 
d(R^{-1}AR+R^{-1}dR)\\
&& +(R^{-1}AR+R^{-1}dR)\ww(R^{-1}AR+R^{-1}dR)\\
&=&(dR^{-1})\ww AR+R^{-1}(dA)R-R^{-1}A\ww dR\\
&&(dR^{-1})\ww(dR)+R^{-1}A\ww AR+R^{-1}A\ww(dR)\\
&&-(dR^{-1})\ww AR-(dR^{-1})\ww(dR)\\
&=&R^{-1}(dA+A\ww A)R
\ees
where we have used $R^{-1}dR+(dR^{-1})R==d(1)=0$ above. 
Thus the PC-form for gauge connection field $A$ should look like 
\be \Theta_A= {\rm Tr}[*F\ww (dA+A\ww A)]-H_A. \ee
where $F$  has the transformation $F\to R^{-1}FR$.

The Hamiltonian  4-form is taken to be
\be H_A=\frac{1}{2}{\rm Tr}(*F)\ww F \ee
which gives
\bes\Xi &=& -d\Theta_A \\
&=& -d[{\rm Tr}*F\ww (dA+A\ww A)]+dH_A\\
&=&-{\rm Tr}[(d*F)\ww (dA+A\ww A)]-{\rm Tr}[*F\ww(dA\ww A-A\ww dA)]\\
&& +  {\rm Tr}(d*F)\ww F
\ees
where we have used the formula 
\bes d(*F\ww F)=2(d*F)\ww F.\ees
Now, we can write
\bes {\rm Tr}[*F\ww dA\ww A]={\rm Tr}[A\ww *F\ww dA]={\rm Tr}[A\ww *F\ww (dA+A\ww A-F)]\ees
where we have added two terms which are actually zero because they carry five
factors of $dt$. Similarly,
\bes {\rm Tr}[*F\ww A\ww dA]={\rm Tr}[*F\ww A\ww (dA+A\ww A-F)].\ees
Thus
\bes -d\Theta_A =-{\rm Tr}[(d*F+A\ww *F-*F\ww A)\ww(dA+A\ww A-F)]\ees
The equations of motion are finally,
\be d*F+A\ww *F-*F\ww A=0,\qquad F=dA+A\ww A \ee

\section{Matter and gauge fields together}

When both the matter fields $\phi$ and gauge fields $A$ are present, the PC-form
is the sum of the two
\be \Theta=\Theta_\phi-H_\phi+\Theta_A-H_A .\ee
We vary both $\phi$ and $A$ fields. The expressions have been
calculated before and the only change now is that the term
$*p\ww dA\phi$ which we threw away in $-d\Theta_\phi$
 when $A$ was an external field depending explicitly
on spacetime, has to be kept and combined with the similar $dA$ term
in $-d\Theta_A$ There is also a minor problem of changing $A$ in the 
$\tau_a$ representation to the adjoint representation.

The concerned term is $(*p)\ww dA^a\tau_a\phi$. From $A=A^aT_a$ we can take the trace 
and infer ${\rm Tr}(AT_b)=A^ak_{ab}$ Therefore, using the inverse matrix,
$A^a=k^{ab}{\rm Tr}(AT_b)$. The combines PC-form is then
\bes \Xi &=& -(d*p+*p\ww A-m^2\phi (*1))\ww(d\phi+A\phi-p)\\
&&+(*p)\tau_a\phi k^{ab}{\rm Tr}[T_b(dA+A\ww A)]\\
&&-{\rm Tr}[(d*F+A\ww *F-*F\ww A)\ww(dA+A\ww A-F)]
\ees
where we have completed $dA$ to $dA+A\ww A-F$ because $A\ww A-F$ has two factors
of $dt$ which will give zero when multiplied to three in $*p$. Thus finally
we can bring this term in the last set of terms
\be \Xi &=& -(d*p+*p\ww A-m^2\phi (*1))\ww(d\phi+A\phi-p)\nonumber \\
&&-{\rm Tr}[(d*F+A\ww *F-*F\ww A-(*p\tau_a\phi)k^{ab}T_b)\nonumber \\
&&\ww(dA+A\ww A-F)]
\ee
The Hamiltonian equations are therefore as follows. The definition of momenta
is given by
\be p &=& d\phi +A\phi,\\
F&=&dA+A\ww A,\ee
and the `field equations' are, finally, 
\be 
d*p+*p\ww A-m^2\phi (*1) &=&0,\\
d*F+A\ww *F-*F\ww A-(*p\tau_a\phi)k^{ab}T_b &=& 0.
\ee

\begin{appendix}
\section{Notation}
We use notation as given for example in \cite{sharan3} or \cite{AMP}.

The spacetime is a Riemannian space with coordinates $t^\mu,\mu=0,1,2,3$.
Basis vectors in a tangent space are written $\ppp_\mu=\ppp/\ppp t^\mu$
The metric is given by the inner product 
$\la \ppp_\mu,\ppp_\nu\ra=g_{\mu\nu}$. The cotangent spaces have
basis elements $dt^\mu$ with $\la dt^\mu,dt^\nu\ra=g^{\mu\nu}$.
The metric has signature $(-1,1,1,1)$. The wedge product is defined so that
$\alpha\ww \beta=\alpha\tp\beta-\beta\tp\alpha$ for one-forms $\alpha$ and $\beta$.
The exterior derivative is defined so that for an $r$-form 
$\alpha=a_{\mu_1\dots\mu_r}dt^{\mu_1}\ww\dots\ww dt^{\mu_r}$ the derivative 
is the $(r+1)$-form
\bes d\alpha=a_{\mu_1\dots\mu_r,\nu}dt^\nu\ww dt^{\mu_1}\ww\dots\ww dt^{\mu_r}.\ees
The Hodge star is a linear operator that maps $r$-forms into $(4-r)$-forms
in our four-dimensional space.
The definition is 
\bes *(dt^{\mu_1}\ww\dots\ww dt^{\mu_r})&=&[(4-r)!]^{-1}\sqrt{-g}g^{\mu_1\nu_1}\dots\\
&&g^{\mu_r\nu_r}\varepsilon_{\nu_1\dots\nu_r\nu_{r+1}\dots\nu_4}dt^{\nu_{r+1}}\dots dt^{\nu_4}\ees
where $g$ denotes the determinant of $g_{\mu\nu}$ and $\varepsilon$ is the antisymmetric tensor
defined with $\varepsilon_{0123}=1$. The one-dimensional space of 0-forms has 
the unit vector equal to real number $1$. The one-dimensional space of 
4-forms has the chosen orientation given by the unit vector 
$\varepsilon=n^0\ww n^1\ww n^2\ww n^3$
where $n^\mu$ are the orthonormal basis vectors in the four-dimensional
space of 1-forms. In the coordinate basis
$\varepsilon=\sqrt{-g}dt^0\ww dt^1\ww dt^2\ww dt^3$. The star operator acting on the zero
form equal to constant number $1$ is denoted by 
$*1=\varepsilon=\sqrt{g}dt^0\ww dt^1\ww dt^2\ww dt^3$.
We have the simple result that $dt^\mu\ww *dt^\nu=-*dt^\nu\ww dt^\mu=g^{\mu\nu}(*1)$

The interior product $i(X)$ of a vector $X$ with an $r$-form $\alpha$ gives 
an $(r-1)$-form $i(X)\alpha$ defined by
\bes (i(X)\alpha)(Y_1,\dots,Y_{r-1})=\alpha(X,Y_1,\dots,Y_{r-1})\ees
When it is more convenient we will denote the interior product operator by $i_X$
in place of $i(X)$.

Two successive applications of interior products on a form are denoted by
\bes i(X,Y)\alpha\equiv [i(X)\circ i(Y)]\alpha = i(X)[i(Y)\alpha] \ees
Note that $i(X,Y)=-i(Y,X)$. Similarly successive applications \\$i(XY\dots Z)$ of many such
interior products can be defined. If $\alpha$ is an $r$-form then
\bes i(X)(\alpha\ww\beta)=[i(X)\alpha]\ww\beta+(-1)^r\alpha\ww i(X)\beta\ees
\end{appendix}

\end{document}